\documentclass[12pt]{article} 
\usepackage{amsmath,amssymb,graphicx}

\makeatletter
\@addtoreset{equation}{section}
\makeatother
\def\theequation{\thesection.\arabic{equation}}
\def\thefootnote{%
\fnsymbol{footnote}}

\newcommand{\dps}{\displaystyle }
\newcommand{\e }{\varepsilon }

\newcommand{\al }{\alpha }
\newcommand{\de }{\delta }

\newcommand{\ket }{\rangle }
\newcommand{\bra }{\langle }
\newcommand{\ga }{\gamma }
\newcommand{\la }{\lambda }
\newcommand{\Hw }{H_{\hbox{w}}} 
\newcommand{\im}{\hbox{Im}}
\newcommand{\re}{\hbox{Re}}
\newcommand{\La }{\Lambda }
\newcommand{\De }{\Delta }
\newcommand{\no }{\nonumber }
\newcommand{\kob }{\overline{K^0}}
\newcommand{\ko }{K^0}
\newcommand{\nub }{\overline{\nu }}

\begin{document}
\begin{flushright}
NUP-A-2000-3\\February 2000
\end{flushright}
~\\ ~\\ 
\begin{center}
\Large{$CP$,~$T$ and/or $CPT$ Violations in the $\ko$-$\kob$ System}\\ 
\large{--Implications of the KTeV,~NA48 and CPLEAR Results--}
\end{center} 
~\\ ~\\ 
\begin{center}
Yutaka Kouchi,
Yoshihiro Takeuchi\footnote{E-mail address: yytake@phys.cst.nihon-u.ac.jp} and
S. Y. Tsai\footnote{E-mail address: tsai@phys.cst.nihon-u.ac.jp}
\end{center}

\begin{center}
{\it Atomic Energy Research Institute and Department of Physics \\ 
College of Science and Technology, Nihon University \\ 
Kanda-Surugadai, Chiyoda-ku, Tokyo 101-8308, Japan }
\end{center}
~\\ ~\\ 

%%%%%%% abstract %%%%%%%%%%%%%%%%

\begin{abstract}
Possible violation of $CP$,~$T$ and/or $CPT$ symmetries in the $\ko$-$\kob$ 
system is studied from a phenomenological point of view. For this purpose, we 
first introduce parameters which represent violation of these symmetries in 
mixing parameters and decay amplitudes in a convenient and well-defined way and, treating these parameters as small, derive formulas which relate them to the 
experimentally measured quantities. We then perform numerical analyses, with the aid of the Bell-Steinberger relation, to derive constraints to these symmetry-violating parameters, firstly paying particular attention to the results 
reported by KTeV Collaboration and NA48 Collaboration, and then with the results reported by CPLEAR Collaboration as well taken into account. A case study, in 
which either $CPT$ symmetry or $T$ symmetry is assumed, is also carried out. It is demonstrated that $CP$ and $T$ symmetries are violated definitively at the 
level of $10^{-4}$ in $2\pi$ decays and presumably at the level of $10^{-3}$ in the $\ko-\kob$ mixing, and that the Bell-Steinberger relation helps us to 
establish $CP$  and $T$ violations being definitively present in the $\ko-\kob$ mixing and to test $CPT$ symmetry to a level of $10^{-4} \sim 10^{-5}$.
\end{abstract}

%%%% S 1: INTORO %%%%
\newpage 

\setcounter{footnote}{0}
\def\thefootnote{%
\mbox{\alph{footnote}}}

\section{Introduction}
Although, on the one hand, all experimental observations up to now are perfectly consistent with $CPT$ symmetry, and, on the other hand, the standard field theory implies that this symmetry should hold exactly, continued experimental, phenomenological and theoretical studies of this and related symmetries are warranted. In this connection, we like to recall, on the one hand, that $CP$ symmetry is violated only at such a tiny level as $10^{-3}$ [1,2], while $CPT$ symmetry is tested at best up to the level one order smaller [3-7] and, on the other hand, that some of the premises of the $CPT$ theorem, e.g., locality, are being challenged by, say, the superstring model.
 
In a series of papers [4-7], we have demonstrated how one may identify or constrain possible violation of $CP$, $T$ and $CPT$ symmetries in the $\ko $-$\kob $ system in a way as phenomenological and comprehensive as possible. For this purpose, we have first introduced parameters which represent violation of these symmetries in mixing parameters and decay amplitudes in a well-defined way and related them to the experimentally measured quantities. We have then carried out numerical analyses, with the aid of the Bell-Steinberger relation [8] and with all the available data on $2\pi$, $3\pi$, $\pi^+\pi^-\ga$ and $\pi\ell\nu_{\ell}$ decays used as inputs, to derive constraints to these symmetry-violating parameters. It has been shown among other things that the new results on the asymptotic leptonic asymmetries obtained by CPLEAR Collaboration [9] allow one for the first time to constrain to some extent possible $CPT$ violation in the $\pi\ell\nu_{\ell}$ decay modes.\footnote{We afterwards became aware that CPLEAR Collaboration themselves [10] had also, by an analysis more or less similar to ours, reached the similar conclusion independently.} 

The present work is a continuation of the previous works, which is new particularly in the following points:\\
(1)~The new results on $\re(\e'/\e)$, etc., from the Fermilab KTeV and CERN NA48 experiments [11,12], along with CPLEAR's new data [13-16] and the latest 
version of the data compiled by Particle Data Group(PDG) [17], are used as 
inputs. \\
(2)~A particular attention is paid to clarify what can be said without recourse to the Bell-Steinberger relation and what can be said with the aid of this 
relation. \\
(3)~A case study with either $CPT$ or $T$ symmetry assumed is also carried out. \\
(4)~The relevant decay amplitudes are parametrized in a convenient form, with 
freedom associated with rephasing of both the initial and final states, as 
discussed explicitly and thoroughly in [4,18,19], taken into account.

The paper is organized as follows. The theoretical framework used to describe 
the $\ko$ - $\kob$ system [20], including the Bell-Steinberger relation, 
is recapitulated in Sec.2 and the experimentally measured quantities related 
to $CP$ violation in decay modes of interest to us are enumerated in Sec.3. We 
then parametrize the mixing parameters and decay amplitudes in a convenient and well-defined 
way and give conditions imposed by $CP$, $T$ and/or $CPT$ symmetries on these 
parameters in Sec.4. In Sec.5, experimentally measured quantities are expressed in terms of the parameters defined, treating them as first order small. In Sec.6, paying particular attention to the data provided by KTeV Collaboration and by 
NA48 Collaboration, a numerical analysis is performed, while, in Sec.7, 
with most of the available experimental data, including those reported by CPLEAR Collaboration, used as inputs, a more comprehensive numerical analysis is 
performed. Sec.8 is devoted to a case study, in which the case with $CPT$ 
symmetry assumed and the case with $T$ symmetry assumed are considered 
separately. The results of the analyses are summarized and some concluding remarks are given in  Sec.9.

%%%% S 2 %%%%
\section{The $\ko$-$\kob$ mixing and the Bell-Steinberger \\
relation}
Let $|\ko\ket$ and $|\kob\ket$ be eigenstates of the strong interaction with strangeness $S=+1$ and $-1$, related to each other by $(CP)$, $(CPT)$ and $T$ operations as [4,18,19,21]
%
%%%% (2.1)
%
\begin{equation}
\left. 
	\begin{array}{cc}
	(CP)|\ko\ket = e^{i\al_K}|\kob\ket ~,& (CPT)|\ko\ket = e^{i\beta _K}|\kob\ket ~,\\ 
	(CP)|\kob\ket = e^{-i\al_K}|\ko\ket ~,& (CPT)|\kob\ket = e^{i\beta_K}|\ko\ket ~,\\ 
	T|\ko\ket = e^{i(\beta_K-\al_K)}|\ko\ket ~,&T|\kob \ket = e^{i(\beta_K+\al_K)}|\kob\ket ~.
	\end{array}
\right. 
\end{equation}
Note here that, given the first two where $\al_K$ and $\beta_K$ are arbitrary real parameters, the rest follow from the assumptions $(CP)T=T(CP)=(CPT)$, $(CP)^2=(CPT)^2=1$, and anti-linearity of $T$ and $(CPT)$.
When the weak interaction $\Hw$ is switched on, the $\ko$ and $\kob$ states decay into other states, generically denoted as $|n\ket$, and get mixed. The time evolution of the arbitrary state 
\[
|\Psi(t)\ket = c_1(t)|K_1\ket + c_2(t)|K_2\ket~,
\]
with
\[
|K_1\ket \equiv |\ko\ket~, \qquad |K_2\ket \equiv |\kob\ket~,
\]
is described by a Schr\"odinger-like equation [20,22] 
\[
i\frac{d}{dt}
|\Psi\ket = \La |\Psi\ket~,
\]
or
%
%%% (2.2)
%
\begin{equation}
i\frac{d}{dt}
\left( 
	\begin{array}{c}
	c_1(t)\\ c_2(t) 
	\end{array}
\right) 
=\La 
\left( 
	\begin{array}{c}
	c_1(t)\\ c_2(t) 
	\end{array}
\right) ~.
\end{equation}
The operator or $2 \times 2$ matrix $\La$ may be written as
%
%%% (2.3)
%
\begin{equation}
\La \equiv M - i\Gamma/2~,
\end{equation}
with $M$ (mass matrix) and $\Gamma$ (decay or width matrix) given, to the second 
order in $\Hw$, by
%
%%% (2.4a,b)
%
\begin{subequations}
\begin{eqnarray}
M_{ij} &\equiv& \bra K_i|M|K_j \ket \no \\
       &=& m_K\de_{ij} + \bra K_i|\Hw|K_j \ket \no \\ 
       & & \quad\qquad + \sum _n P \frac{\bra K_i|\Hw|n\ket \bra n|\Hw|K_j \ket}{m_K-E_n}~, \\
\Gamma_{ij} &\equiv& \bra K_i|\Gamma|K_j \ket \no \\
            &=& 2\pi \sum _n \bra K_i|\Hw|n\ket \bra n|\Hw|K_j \ket \de(m_K-E_n)~,\no \\
\end{eqnarray}
\end{subequations}
where the operator $P$ projects out the principal value. The two eigenstates of $\La$ and their respective eigenvalue may be written as 
%
%%% (2.5a,b) 
%
\begin{subequations}
\begin{eqnarray}
|K_S\ket &=& \frac{1}{\sqrt{|p_S|^2+|q_S|^2}}\left( p_S|\ko \ket +q_S|\kob \ket \right) ~, \\
|K_L\ket &=& \frac{1}{\sqrt{|p_L|^2+|q_L|^2}}\left( p_L|\ko \ket -q_L|\kob \ket \right) ~; 
\end{eqnarray}
\end{subequations}
%
%%% (2.6a,b)
%
\begin{subequations}
\begin{eqnarray}
\la _S=m_S-i\frac{\ga _S}{2}~, \\
\la _L=m_L-i\frac{\ga _L}{2}~.
\end{eqnarray}
\end{subequations}
$m_{S, L}=\re (\la _{S, L})$ and $\ga _{S, L}=-2\im (\la _{S, L})$ are the mass
 and the total decay width of the $K_{S, L}$ state respectively. By definition,
 $\ga_S > \ga_L$ or $\tau_S < \tau_L$ ($\tau_{S,L} \equiv 1/\ga_{S,L}$), and 
the suffices $S$ and $L$ stand for "short-lived" and "long-lived" respectively. The eigenvalues $\la_{S,L}$ and the ratios of the mixing parameters 
$q_{S,L}/p_{S,L}$ are related to the elements of the mass-width matrix $\La$ as 
%
%%% (2.7),(2.8)
%
\begin{equation}
\la_{S,L} = \pm E + (\La_{11} + \La_{22})/2~,
\end{equation}
\begin{equation}
q_{S,L}/p_{S,L} = \La_{21}/[E \pm (\La_{11} - \La_{22})/2]~,
\end{equation}
where
%
%%% (2.9)
%
\begin{equation}
E \equiv [\La_{12}\La_{21} + (\La_{11} - \La_{22})^2/4]^{1/2}.
\end{equation}

>From the eigenvalue equation of $\La$, one may readily derive the well-known Bell-Steinberger relation [8]:
%
%%% (2.10)
%
\begin{equation}
\left[ \frac{\ga_S+\ga_L}{2}-i(m_S-m_L) \right] \bra K_S|K_L\ket = \bra K_S|\Gamma |K_L\ket ~,
\end{equation}
where 
%
%%% (2.11)
%
\begin{equation}
\bra K_S|\Gamma |K_L\ket = 2\pi \sum _n\bra K_S|\Hw|n\ket \bra n|\Hw |K_L\ket \de (m_K-E_n)~.
\end{equation}
%
%%%% S 3 %%%%
\section{Decay modes}
\label{sec:3}
The $\ko$ and $\kob$ (or $K_S$ and $K_L$) states have many decay modes, among which we are interested in $2\pi$, $3\pi$, $\pi^+\pi^-\ga$ and semileptonic modes.
%
%%%% S 3.1 %%%%
\subsection{$2\pi$ modes }
\label{sec:3.1}
The experimentally measured quantities related to $CP$ violation are $\eta _{+-}$ and $\eta _{00}$ defined by 
%
%%% (3.1a,b)
%
\begin{subequations}
\begin{equation}
\eta _{+-} \equiv |\eta _{+-}|e^{i\phi _{+-}} \equiv \frac{\bra \pi ^+\pi ^-,\mbox{outgoing}|\Hw |K_L\ket }{\bra \pi ^+\pi ^-,\mbox{outgoing}|\Hw |K_S\ket }~, 
\end{equation}
\begin{equation}
\eta _{00} \equiv |\eta _{00}|e^{i\phi _{00}} \equiv \frac{\bra \pi ^0\pi ^0,\mbox{outgoing}|\Hw |K_L\ket }{\bra \pi ^0\pi ^0,\mbox{outgoing}|\Hw |K_S\ket }~. 
\end{equation}
\end{subequations}
Defining
%
%%% (3.2)
%
\begin{equation}
\omega \equiv \frac{\bra (2\pi )_2|\Hw |K_S\ket }{\bra (2\pi )_0|\Hw |K_S\ket }~,
\end{equation}
%
%%% (3.3)
%
\begin{equation}
\eta _I \equiv |\eta_I|e^{i\phi_I} \equiv \frac{\bra (2\pi )_I|\Hw |K_L\ket }{\bra (2\pi )_I|\Hw |K_S\ket }~,
\end{equation}
where $I$=1 or 2 stands for the isospin of the $2\pi$ states, one gets
%
%%% (3.4a,b)
%
\begin{subequations}
\begin{eqnarray}
\eta _{+-} &=& \frac{\eta _0+\eta_2\omega '}{1+\omega '}~, \\
\eta _{00} &=& \frac{\eta _0-2\eta _2\omega '}{1-2\omega '}~,
\end{eqnarray}
\end{subequations}
where 
%
%%% (3.5)
%
\begin{equation}
\omega ' \equiv \frac{1}{\sqrt{2}}\omega e^{i(\de _2-\de _0)}~,
\end{equation}
$\de _I$ being the S-wave $\pi \pi $ scattering phase shift for the isospin $I$ state at an energy of the rest mass of $\ko $. $\omega $ is a measure of deviation from the $\De I=1/2$ rule, and may be inferred, for example, from 
%
%%% (3.6)
%
\begin{eqnarray}
r&\equiv &\frac{\ga _S(\pi ^+\pi ^-)-2\ga _S(\pi ^0\pi ^0)}{\ga _S(\pi ^+\pi ^-)+\ga _S(\pi ^0\pi ^0)} \no \\ 
&=& \frac{4\re (\omega ')-2|\omega '|^2}{1+2|\omega '|^2} ~.
\end{eqnarray}
Here and in the following, $\ga _{S,L}(n)$ denotes the partial width for $K_{S,L}$ to decay into the final state $|n\ket $.

%%%% S 3.2 %%%%
\subsection{$3\pi$ and $\pi^+\pi^-\ga$ modes }
\label{sec:3.2}
The experimentally measured quantities are 
%
%%% (3.7a,b)
%
\begin{subequations}
\begin{equation}
\eta _{+-0}= \frac{\bra \pi ^+\pi ^-\pi ^0,\mbox{outgoing}|\Hw |K_S\ket }{\bra \pi ^+\pi ^-\pi ^0,\mbox{outgoing}|\Hw |K_L\ket }~, 
\end{equation}
\begin{equation}
\eta _{000}= \frac{\bra \pi ^0\pi ^0\pi ^0,\mbox{outgoing}|\Hw |K_S\ket }{\bra \pi ^0\pi ^0\pi ^0,\mbox{outgoing}|\Hw |K_L\ket }~,
\end{equation}
\end{subequations}
%
%%% (3.8)
%
\begin{equation}
\eta _{+-\ga }= \frac{\bra \pi ^+\pi ^-\ga ,\mbox{outgoing}|\Hw |K_L\ket }{\bra \pi ^+\pi ^-\ga ,\mbox{outgoing}|\Hw |K_S\ket }~.
\end{equation}
We shall treat the $3\pi $ $(\pi ^+\pi ^-\ga )$ states as purely $CP$-odd ($CP$-even). 
%%%% S 3.3 %%%%
\subsection{Semileptonic modes}
\label{sec:3.3}
The well measured time-independent asymmetry parameter related to $CP$ violation in semi-leptonic decay modes is
%
%%% (3.9)
%
\begin{equation}
d^{\ell }_L = \frac{\ga _L(\pi ^-\ell ^+\nu _{\ell })-\ga _L(\pi ^+\ell ^-\nub _{\ell })}{\ga _L(\pi ^-\ell ^+\nu _{\ell })+\ga _L(\pi ^+\ell ^-\nub _{\ell })}~,
\end{equation}
where $\ell=e$ or $\mu$. CPLEAR Collaboration [9,14-16] have furthermore for the first time measured two kinds of time-dependent asymmetry parameters
%
%%% (3.10a,b)
%
\begin{subequations}
\begin{equation}
d^{\ell}_1(t)=\frac{|\bra \ell^+|\Hw|\kob(t)\ket|^2-|\bra \ell^-|\Hw|\ko(t)\ket|^2}{|\bra \ell^+|\Hw|\kob(t)\ket|^2+|\bra \ell^-|\Hw|\ko(t)\ket|^2}~,
\end{equation}
\begin{equation}
d^{\ell}_2(t)=\frac{|\bra \ell^-|\Hw|\kob(t)\ket|^2-|\bra \ell^+|\Hw|\ko(t)\ket|^2}{|\bra \ell^-|\Hw|\kob(t)\ket|^2+|\bra \ell^+|\Hw|\ko(t)\ket|^2}~,
\end{equation}
\end{subequations}
where $|\ell^+\ket =|\pi^-\ell^+\nu_{\ell}\ket $ and $|\ell^-\ket =|\pi^+\ell^- \nub_{\ell}\ket$.
%
%%%% Sec. 4 %%%%
%
\section{Parametrization and conditions imposed by $CP$, $T$ and $CPT$ symmetries}
We shall parametrize the ratios of the mixing parameters $q_S/p_S$ and $q_L/p_L$ as
%
%%% (4.1)
%
\begin{equation}
\left. 
	\begin{array}{c}
	\frac{\dps{q_S}}{\dps{p_S}}=e^{i\al _K}\frac{\dps{1-\e _S}}{\dps{1+\e _S}}~,\\ \\ 
	\frac{\dps{q_L}}{\dps{p_L}}=e^{i\al _K}\frac{\dps{1-\e _L}}{\dps{1+\e _L}}~,
	\end{array}
\right. 
\end{equation}
and $\e _{S,L}$ further as 
%
%%%% (4.2)
%
\begin{equation}
\e _{S,L}=\e \pm \de ~.
\end{equation}
>From Eqs.(2.7), (2.8) and (2.9), treating $\e$ and $\de$ as small parameters, one may derive [4]
%
%%% (4.3a,b)
%
\begin{subequations}
\begin{eqnarray}
\De m &\simeq& 2\re(M_{12}e^{i\alpha_K})~, \\
\De \ga &\simeq& 2\re(\Gamma_{12}e^{i\alpha_K})~,
\end{eqnarray}
\end{subequations}
%
%%% (4.4a,b)
%
\begin{subequations}
\begin{eqnarray}
\e &\simeq& (\La_{12}e^{i\alpha_K}-\La_{21}e^{-i\alpha_K})/2\De \la~, \\
\de &\simeq& (\La_{11}-\La_{22})/2\De \la~,
\end{eqnarray}
\end{subequations}
from which it follows that [4,5]
%
%%% (4.5a,b)
%
{
\setcounter{enumi}{\value{equation}}
\addtocounter{enumi}{1}
\setcounter{equation}{0}
\renewcommand{\theequation}{\thesection.\theenumi\alph{equation}}
\begin{equation}
\e _{\| }\equiv \re [\e \exp (-i\phi _{SW})]\simeq \frac{-2\im (M_{12}e^{i\al _K})}{\sqrt{(\ga _S-\ga _L)^2+4(\De m)^2}}~,
\end{equation}
\begin{equation}
\e _{\perp }\equiv \im [\e \exp (-i\phi _{SW})]\simeq \frac{\im (\Gamma _{12}e^{i\al _K})}{\sqrt{(\ga _S-\ga _L)^2+4(\De m)^2}}~,
\end{equation}
\setcounter{equation}{\value{enumi}}}%
%
%%% (4.6a,b) 
%
{
\setcounter{enumi}{\value{equation}}
\addtocounter{enumi}{1}
\setcounter{equation}{0}
\renewcommand{\theequation}{\thesection.\theenumi\alph{equation}}
\begin{equation}
\de _{\|}\equiv \re [\de \exp (-i\phi _{SW})]\simeq \frac{(\Gamma _{11}-\Gamma _{22})}{2\sqrt{(\ga _S-\ga _L)^2+4(\De m)^2}}~,
\end{equation}
\begin{equation}
\de _{\perp }\equiv \im [\de \exp (-i\phi _{SW})]\simeq \frac{(M_{11}-M_{22})}{\sqrt{(\ga _S-\ga _L)^2+4(\De m)^2}}~,
\end{equation}
\setcounter{equation}{\value{enumi}}}%
where 
%
%%% (4.7a,b) 
%
\begin{subequations}
\begin{eqnarray}
\De m \equiv m_S-m_L~,~~  \De \ga &\equiv& \ga_S-\ga_L~, ~~ \De \la \equiv \la_S-\la_L~, \\ 
\phi _{SW} &\equiv& \tan ^{-1}\left( \frac{-2\De m}{\De \ga} \right)~. 
\end{eqnarray}
\end{subequations}
$\phi _{SW}$ is often called the superweak phase. 

Paying particular attention to the $2\pi$ and semileptonic decay modes, we shall  parametrize amplitudes for $\ko$ and $\kob$ to decay into $|(2\pi)_I\ket$ as
%
%%% (4.8a,b)
%
\begin{subequations}
\begin{eqnarray}
	\bra(2\pi)_I|\Hw|\ko\ket &=& F_I(1+\e_I)e^{i\al_K/2}~, \\
	\bra(2\pi)_I|\Hw|\kob\ket &=& F_I(1-\e_I)e^{-i\al_K/2}~,
\end{eqnarray}
\end{subequations}
and amplitudes for $\ko$ and $\kob$ to decay into $|\ell^+\ket$ and $|\ell^-
\ket$ as
%
%%% (4.9a,b,c,d)
%
\begin{subequations}
\begin{eqnarray}
	\bra\ell^+|\Hw|\ko\ket &=& F_{\ell}(1+\e_{\ell})e^{i\al_K/2}~, \\ 
	\bra\ell^-|\Hw|\kob\ket &=& F_{\ell}(1-\e_{\ell})e^{-i\al_K/2}~, \\ 
	\bra\ell^+|\Hw|\kob\ket &=& x_{\ell+}F_{\ell}(1+\e_{\ell})e^{-i\al_K/2}~, \\
	\bra\ell^-|\Hw|\ko\ket &=& x_{\ell-}^*F_{\ell}(1-\e_{\ell})e^{i\al_K/2}~.
\end{eqnarray}
\end{subequations}
$x_{\ell+}$ and $x_{\ell-}$, which measure violation of the $\De S=\De Q$ rule, will further be parametrized as
%
%%% (4.10)
%
\begin{equation}
x_{\ell+} = x_{\ell}^{(+)} + x_{\ell}^{(-)}~, \qquad x_{\ell-} = x_{\ell}^{(+)} - x_{\ell}^{(-)}~.
\end{equation}

Our amplitude parameters $F_I$, $\e_I$, $F_{\ell}$, $\e_{\ell}$, $x_{\ell}^{(+)}$ and $x_{\ell}^{(-)}$, and our mixing parameters $\e$ and $\de$ as well, are all invariant with respect to rephasing of $|\ko\ket$ and $|\kob\ket$,
%
%%% (4.11)
%
\begin{equation}
|\ko \ket \to |\ko \ket '=|\ko \ket e^{-i\xi _K}~, ~~ |\kob \ket \to |\kob \ket '=|\kob \ket e^{i\xi _K}~,
\end{equation}
in spite that $\al_K$ itself is not invariant with respect to this rephasing [4,18]. $F_I$, $\e_I$, $F_{\ell}$ and $\e_{\ell}$ are however not invariant with respect to rephasing of the final states [7,19],
%
%%% (4.12a,b) 
%
\begin{subequations}
\begin{eqnarray}
	&|(2\pi )_I\ket \to |(2\pi )_I\ket '=|(2\pi )_I\ket e^{-i\xi_I}~, \\
	&|\ell^+\ket \to |\ell^+\ket '=|\ell^+\ket e^{-i\xi_{\ell+}}~, ~~ 
	 |\ell^-\ket \to |\ell^-\ket '=|\ell^-\ket e^{-i\xi_{\ell-}}~,
\end{eqnarray}
\end{subequations}
nor are the relative $CP$ and $CPT$ phases $\al_{\ell}$, $\beta_I$ and $\beta_{\ell}$ defined in such a way as
%
%%% (4.13)
%
\begin{subequations}
\begin{eqnarray}
  &CP|\ell^+\ket = e^{i\al_{\ell}}|\ell^-\ket~, \\
  &CPT|(2\pi)_I\ket = e^{i\beta_I}|(2\pi)_I\ket~, ~~ 
   CPT|\ell^+\ket = e^{i\beta_{\ell}}|\ell^-\ket~.
\end{eqnarray}
\end{subequations}
One may convince himself [4,18,19] that freedom associated with choice of 
$\xi_I$, $\xi_{\ell+} + \xi_{\ell-}$ and $\xi_{\ell+} - \xi_{\ell-}$ allows one, without loss of generality, to take\footnote{Note that, although freedom 
associated with $\xi_K$ and $\xi_{\ell+} - \xi_{\ell-}$ allows one to take 
$\al_K=0$ and $\al_{\ell} = 0$ (instead of $\im(\e_{\ell})=0$) respectively, we  prefer not to do so. Note also that our parametrization (4.9a,b) is similar to, but different from the one more widely adopted [23,24],
\[
\bra\ell^+|\Hw|\ko\ket = F_{\ell}(1-y_{\ell})~, \qquad
\bra\ell^-|\Hw|\kob\ket = F_{\ell}^*(1+y_{\ell}^*)~,
\]
and that, nevertheless, our $\re(\e_{\ell})$ is exactly equivalent to 
$-\re(y_{\ell})$ introduced through these equations and also to $-\re(y)$ 
defined in [14-16].}
%
%%% (4.14)
%
\begin{equation}
\im(F_I) = 0~, \quad \im(F_{\ell}) = 0~, \quad \im(\e_{\ell}) = 0~,
\end{equation}
respectively, and that $CP$, $T$ and $CPT$ symmetries impose such conditions as
%
%%% (4.15)
%
\begin{equation}
\left. 
	\begin{array}{ccl}
	CP~\mbox{symmetry} &:& \e = 0,~\de = 0,~\e_I = 0,~\re(\e_{\ell}) = 0,~\\
		&&\im(x_{\ell}^{(+)}) = 0,~\re(x_{\ell}^{(-)}) = 0~; \\ 
	T~\mbox{symmetry} &:& \e = 0,~\im(\e_I) = 0,~\im(x_{\ell}^{(+)}) = 0,~\\
		&&\im(x_{\ell}^{(-)}) = 0~; \\ 
	CPT~\mbox{symmetry} &:& \de = 0,~\re(\e_I) = 0,~\re(\e_{\ell}) = 0,~\\
		&&\re(x_{\ell}^{(-)}) = 0,~\im(x_{\ell}^{(-)}) = 0~. 
	\end{array}
\right. 
\end{equation}
Among these parameters, $\e$ and $\de$ will be referred to as indirect parameters and the rest as direct parameters.\footnote{As emphasized in [18], classification of the symmetry-violating parameters into "direct" and "indirect" ones makes sense only when they are defined in such a way that they are invariant under rephasing of $|\ko\ket$ and $|\kob\ket$, Eq.(4.11).}

%%%% S 5 %%%%
\section{Formulas relevant for numerical analysis }

We shall adopt a phase convention which gives Eq.(4.14). Observed or expected smallness of violation of $CP$, $T$ and $CPT$ symmetries and of the $\De I = 1/2$ and $\De Q = \De S$ rules allows us to treat all our parameters, $\e$, $\de$, $\e_I$ , $\e_{\ell}$, $x_{\ell}^{(+)}$, $x_{\ell}^{(-)}$ as well as $\omega'$ as small, \footnote{As a matter of fact, we have already assumed that $CP$, $T$ and $CPT$ violations are small in deriving Eqs.(4.3a) $ \sim $ (4.6b) and in parametrizing the relevant amplitudes as in Eqs.(4.8) and (4.9).} and, from Eqs.(3.2), (3.3), (3.4a,b), (3.6), (3.9) and (3.10a,b), one finds, to the leading order in these small parameters,
%
%%% (5.1)
%
\begin{equation}
\omega \simeq \re(F_2)/\re(F_0)~,
\end{equation}
%
%%% (5.2)
%
\begin{equation}
\eta _I\simeq \e -\de +\e_I~,
\end{equation}
%
%%% (5.3a,b)
%
\begin{subequations}
\begin{eqnarray}
\eta_{+-} &\simeq& \eta_0 + \e'~, \\
\eta_{00} &\simeq& \eta_0 - 2\e'~,
\end{eqnarray}
\end{subequations}
%
%%% (5.4)
%
\begin{equation}
r \simeq 4\re(\omega')~,
\end{equation}
%
%%% (5.5)
%
\begin{equation}
d_L^{\ell} \simeq 2\re(\e-\de)+2\re(\e_{\ell}-x_{\ell}^{(-)})~,
\end{equation}
%
%%% (5.6a,b) 
%
\begin{subequations}
\begin{eqnarray}
d_1^{\ell}(t \gg \tau_S) \simeq 4\re(\e)+2\re(\e_{\ell}-x_{\ell}^{(-)})~, \\
d_2^{\ell}(t \gg \tau_S) \simeq 4\re(\de)-2\re(\e_{\ell}-x_{\ell}^{(-)})~,
\end{eqnarray}
\end{subequations}
where
%
%%% (5.7) 
%
\begin{equation}
\e' \equiv (\eta_2-\eta_0)\omega'~.
\end{equation}
Note that $d_L^{\ell}$, $d_1^{\ell}(t \gg \tau_S)$ and $d_2^{\ell}(t \gg \tau _S)$ are not independent:
%
%%% (5.8)
%
\begin{equation}
d_L^{\ell} \simeq [~d_1^{\ell}(t \gg \tau_S) - d_2^{\ell}(t \gg \tau_S)~]/2~.
\end{equation}

>From Eqs.(5.3a,b), it follows that
%
%%% (5.9)
%
\begin{equation}
\eta_0 \simeq (2/3)\eta_{+-}(1+(1/2)|\eta_{00}/\eta_{+-}|e^{i\De\phi})~,
\end{equation}
and, treating $|\e'/\eta_0|$ as a small quantity, which is justifiable empirically (see below), one further obtains
%
%%% (5.10)
%
\begin{equation}
\eta _{00}/\eta _{+-} \simeq 1 - 3\e'/\eta_0~,
\end{equation}
or
%
%%% (5.11a,b)
%
\begin{subequations}
\begin{eqnarray}
\re(\e'/\eta_0) &\simeq& (1/6)(1 - |\eta_{00}/\eta_{+-}|^2)~, \\
\im(\e'/\eta_0) &\simeq& -(1/3)\De\phi~,
\end{eqnarray}
\end{subequations}
where
%
%%% (5.12)
%
\begin{equation}
\De\phi \equiv \phi_{00} - \phi_{+-}~.
\end{equation}
On the other hand, from Eqs.(3.5), (5.1), (5.2) and (5.7), one may derive
%
%%% (5.13)
%
\begin{equation}
\e'/\eta_0 = -i\re(\omega')(\e_2-\e_0)e^{-i\De\phi'}/[|\eta_0|\cos(\de_2-\de_0)]~,
\end{equation}
where
%
%%% (5.14)
%
\begin{equation}
\De\phi' \equiv \phi_0 - \de_2 + \de_0 - \pi/2~.
\end{equation}

Furthermore, noting that 
\[
\bra K_S|K_L\ket \simeq 2[\re(\e) - i\im(\de)]~,
\]
one may use the Bell-Steinberger relation, Eq.(2.10), to express $\re(\e)$ and $\im(\de)$ in terms of measured quantities. By taking $2\pi $, $3\pi $, $\pi ^+\pi ^-\ga $ and $\pi \ell \nu_{\ell }$ intermediate states into account in Eq.(2.11) and making use of the fact $\ga _S \gg \ga _L$, we derive
%
%%% (5.15)
%
\begin{eqnarray}
\re (\e ) & \simeq & \frac{1}{\sqrt{\ga _S^2+4(\De m)^2}}\times \no \\ 
 &&  \Big{[} ~\ga _S(\pi ^+\pi ^-)|\eta _{+-}|\cos (\phi _{+-}-\phi _{SW}) \no \\ 
 && ~~+\ga _S(\pi ^0\pi ^0)|\eta _{00}|\cos (\phi _{00}-\phi _{SW}) \no \\ 
 && ~~+\ga _S(\pi ^+\pi ^-\ga )|\eta _{+-\ga }|\cos (\phi _{+-\ga }-\phi _{SW}) \no \\ 
 && ~~+\ga _L(\pi ^+\pi ^-\pi ^0)\{ \re (\eta _{+-0})\cos \phi _{SW} -\im (\eta _{+-0})\sin \phi _{SW}\} \no \\ 
 && ~~+\ga _L(\pi ^0\pi ^0\pi ^0)\{ \re (\eta _{000})\cos \phi _{SW} -\im (\eta _{000})\sin \phi _{SW}\} \no \\ 
 && ~~+2\sum _{\ell }\ga _L(\pi \ell \nu _{\ell })\{ \re(\e_{\ell})\cos \phi _{SW} -\im(x_{\ell}^{(+)})\sin \phi _{SW}\} \Big{]}, 
\end{eqnarray}
%
%%% (5.16)
%
\begin{eqnarray}
\im (\de ) & \simeq & \frac{1}{\sqrt{\ga _S^2+4(\De m)^2}}\times \no \\ 
 &&  \Big{[} ~-\ga _S(\pi ^+\pi ^-)|\eta _{+-}|\sin (\phi _{+-}-\phi _{SW}) \no \\ 
 && ~~-\ga _S(\pi ^0\pi ^0)|\eta _{00}|\sin (\phi _{00}-\phi _{SW}) \no \\ 
 && ~~-\ga _S(\pi ^+\pi ^-\ga )|\eta _{+-\ga }|\sin (\phi _{+-\ga }-\phi _{SW}) \no \\ 
 && ~~+\ga _L(\pi ^+\pi ^-\pi ^0)\{ \re (\eta _{+-0})\sin \phi _{SW} +\im (\eta _{+-0})\cos \phi _{SW}\} \no \\ 
 && ~~+\ga _L(\pi ^0\pi ^0\pi ^0)\{ \re (\eta _{000})\sin \phi _{SW} +\im (\eta _{000})\cos \phi _{SW}\} \no \\ 
 && ~~+2\sum _{\ell }\ga _L(\pi \ell \nu _{\ell })\{ \re (\e_{\ell })\sin \phi _{SW} +\im(x_{\ell}^{(+)})\cos \phi _{SW}\} \Big{]}.
\end{eqnarray}
If, however, one retains the contribution of the $2\pi$ intermediate states 
alone, which is justfiable empirically, the Bell-Steinberger relation gives 
simply
%
%%% (5.17)
%
\begin{equation}
\re(\e) - i\im(\de) \simeq |\eta_0|e^{i\De\phi"}\cos\phi_{SW}~,
\end{equation}
where
%
%%% (5.18)
%
\begin{equation}
\De\phi" \equiv \phi_0 - \phi_{SW}~.
\end{equation}
It is to be noted that exactly the same equations as Eq.(5.17) can be derived from Eqs.(4.5b) and (4.6a).%

%%%% Sec.6 %%%%
\section{Numerical analysis (1) -Constraints from the \\
KTeV and NA48 data-}
The data used as inputs in the numerical analysis given below are 
tabulated in Table 1. As the value of $|\eta_{00}/\eta_{+-}|$ or 
$\re(\e'/\eta_0)$, we adopt those [11,12] reported by KTeV Collaboration and 
NA48 Collaboration,\footnote{Note that $\eta_0$ corresponds to $\e$ used in [11,12]. Since only $\re(\e'/\e)$, but not $|\eta_{00}/\eta_{+-}|^2$, is 
reported  explicitly in [11], we take a weighted average of the two values of 
$\re(\e'/\e)$ reported in [11,12] and list this in Table 2 below.} and as 
the values of $\De m$, $\tau_S$ and $\De\phi$, we use those  reported by KTeV 
Collaboration[11]. As for $\de_2-\de_0$, we use $(-42\pm20)^{\circ}$, i.e., the  Chell-Olsson value [25] with the error arbitrarily extended by a factor of five to take account of its possible uncertainty [26]. All the other data are from 
Particle Data Group (PDG) [17].

Our analysis consists of two parts:

{\it The first half.}~~We use Eq.(4.7b) to find $\phi_{SW}$ from $\De m$ 
and $\ga_S$, use Eqs.(3.6) and (5.4) to find $\re(\omega')$ from 
$\ga_S(\pi^+\pi^-)/\ga_S$ and $\ga_S(\pi^0\pi^0)/\ga_S$, and 
further use Eqs.(5.11a,b) and (5.9) to find 
$\re(\e'/\eta_0)$, $\im(\e'/\eta_0)$, $|\eta_0|$ and $\phi_0$ from 
$|\eta_{00}/\eta_{+-}|$, $\De \phi$, $|\eta_{+-}|$ and $\phi_{+-}$. These 
results are shown as the intermediate outputs in Table 2.

{\it The second half.}~~The values of $\eta_0$, $\e'/\eta_0$, $\phi_{SW}$ 
and $\re(\omega')$ obtained, supplemented with the value of $\de_2-\de_0$, are 
used as inputs to find $\re(\e_2-\e_0)$ and $\im(\e_2-\e_0)$ with the help of 
Eqs.(5.13) and (5.14), and to find $\re(\e)$ and $\im(\de)$ with the help of 
Eq.(5.17).  The values of $\re(\e)$ and $\im(\de)$ are 
in turn used to constrain $\re(\de-\e_0)$ and $\im(\e+\e_0)$ through Eq.(5.2) 
and constrain $\re(\de-\e_{\ell}+x_{\ell}^{(-)})$ through Eq.(5.5). The 
numerical results obtained are shown in Table 3. The value of $\re(\e-\de+
\e_{\ell}-x_{\ell}^{(-)})$, which is nothing but the value of $d_L^{\ell}/2$, is 
also shown.
%
%%%%%% table 1.
% For tables use
\begin{table}

\caption{Input data (1).}
\label{tab:1}       % Give a unique label
% For LaTeX tables use
 \renewcommand{\footnoterule}{}
 \begin{minipage}{15cm}
	\begin{center}
   \begin{tabular}{c|c|c|c|l}
    \hline \hline 
    &Quantity & Value  &Unit & Ref. \\ \hline 
    
    &$\tau _S$ &$0.896\pm 0.0007$ &$10^{-10}s$ & [11] \\ 
    &$\tau _L$ &$5.17\pm 0.04$ &$10^{-8}s$ & [17] \\ 
    &$-\De $ &$0.5268\pm 0.0015$ &$10^{10}s^{-1}$ & [11] \\ \hline 
    
    $2\pi $& $\ga _S(\pi ^+\pi ^-)/\ga _S$ & $68.61\pm 0.28$ & $\% $ & [17] \\ 
    &$\ga _S(\pi ^0\pi ^0)/\ga _S$ & $31.39\pm 0.28$ & \% & [17] \\ 
    &$\de_2-\de_0$ &$(-42\pm 20)$ & $^\circ$ & [25]
    \footnote{Error extended arbitrarily by a factor of five.} \\ 
    &$|\eta _{+-}|$ &$2.285\pm 0.019$ &$10^{-3}$ & [17] \\ 
    &$\phi _{+-}$ &$43.5\pm 0.6$ &$ ^\circ $ & [17] \\ 
    &$|\eta _{00}/\eta_{+-}|^2$ &$0.9832\pm 0.0025$ &&
    [11] \\ 
    & &$0.9889\pm 0.0044$ &&
    [12] \\ 
    &$\Delta\phi$ &$0.09\pm 0.46$ &$ ^\circ $ & [11] \\ \hline 
    
    $\pi \ell \nu $ &$d^{\ell }_L$ &$3.27\pm 0.12$ &$10^{-3}$ & [17] \\ \hline 
   
   \end{tabular}
	 \end{center}
 \end{minipage}
% Or use
%\vspace*{5cm}  % with the correct table height

\end{table}
%
%%%%%% table 2.

% For tables use
\begin{table}
\caption{Intermediate outputs.}
\label{tab:2}       % Give a unique label
% For LaTeX tables use
	\begin{center}
		\begin{tabular}{c|c|c}
		\hline \hline 
		 Quantity & Value &Unit \\ \hline 

		 $\phi _{SW} $ & $43.40\pm 0.09$ & $^\circ$ \\ 
		 $\re(\omega') $ &$1.458\pm 0.157$ & $\times 10^{-2}$ \\ 
		 $\re(\e'/\eta_0) $ &$2.59\pm 0.36$ & $\times 10^{-3}$ \\ 
		 $\im(\e'/\eta_0) $ &$-0.524\pm 2.676$ & $\times 10^{-3}$ \\ 
		 $|\eta_0| $ &$2.279\pm 0.019$ & $\times 10^{-3}$ \\ 
		 $\phi_0 $ &$43.53\pm 0.94$ & $^\circ$ \\ \hline 

		\end{tabular}
\end{center}
% Or use
%\vspace*{5cm}  % with the correct table height
\end{table}

%
%%%%%% table 3.

% For tables use
\begin{table}
\caption{Constraints (in unit of $10^{-3}$) to $CP$,~$T$ and/or $CPT$-violating prameters (1).}
\label{tab:3}       % Give a unique label
% For LaTeX tables use
	\begin{center}
		\begin{tabular}{c|c}
		\hline \hline 
		 Quantity & Result \\ \hline 

		$\re (\e_2-\e_0)$ &$0.084\pm 0.328$  \\ 
		$\im (\e_2-\e_0)$ &$0.295\pm 0.113$  \\ 
		$\re(\e-\de+\e_{\ell}-x_{\ell}^{(-)})$ & $1.635 \pm 0.060$ \\ \hline 
		$\re (\e)$ & $1.656\pm 0.014$  \\ 
		$\im (\e+ \e_0)$ & $1.566\pm 0.014$  \\ \hline 

		$\im (\de)$ & $-0.004\pm 0.027$  \\  
		$\re (\de -\e_0)$ & $0.004\pm 0.026$  \\

		$\re (\de- \e_{\ell }+x_{\ell }^{(-)})$ & $0.021\pm 0.062$  \\  \hline 
		
		\end{tabular}
\end{center}
% Or use
%\vspace*{5cm}  % with the correct table height
\end{table}

%%%% Sec.7 %%%%
\section{Numerical analysis (2) -Constraints from the \\
CPLEAR results-}

Immediately after CPLEAR Collaboration reported [9] their preliminary 
result on the asymptotic leptonic asymmetries, $d_{1,2}^{\ell}(t \gg \tau_S)$, 
we showed [6] that this result, combined with the other relevant data available, could be used with the help of the Bell-Steinberger relation to constrain many of the $CP$,~$T$ and/or $CPT$-violating parameters introduced. The analysis went as follows. Assuming $\re(x_{\ell}^{(-)})=0$,\footnote{In most of the 
experimental analyses prior to those [15,16] by CPLEAR Collaboration, either 
$CPT$ symmetry is taken as granted or no distinction is made between $x_{\ell+}$ and $x_{\ell-}$, which implies that $x_{\ell}^{(-)}$ is presupposed to be zero implicitly. Accordingly, we identified $x$ used in [17] with our $x_{\ell}^{(+)}$.} Eqs.(5.6a) and (5.15) were used to find the values of $\re(\e)$ and 
$\re(\e_{\ell})$. The value of $\re(\e_{\ell})$ was then used to constrain 
$\re(\de)$ and $\im(\de)$ through Eqs.(5.6b) and (5.16) respectively and all 
these values were combined with the value of $\eta_0$ to determine or constrain  $\im(\e+\e_0)$ and $\re(\e_0)$.

In order to appreciate the results obtained under the $2\pi$ dominance and 
to separately constrain, as far as possible, the parameters not yet separately 
constrained in the previous section, we now proceed to perform an analysis 
similar to the one explained above [6], with the new results [13-16] reported by CPLEAR Collaboration taken into account. In [14-16], CPLEAR Collaboration have  defined two kinds of experimental asymmetries  $A_T^{exp}(t)$ and $A_{\de}^{exp}(t)$ which are related to $d_{1,2}^{\ell}(t)$ and behave as
%
%%% (7.1a,b)
%
\begin{subequations}
\begin{eqnarray}
A_T^{exp}(t \gg \tau_S) &\simeq& 4\re(\e+\e_{\ell}-x_{\ell}^{(-)})~, \\
A_{\de}^{exp}(t \gg \tau_S) &\simeq& 8\re(\de)~,
\end{eqnarray}
\end{subequations}
and, by performing \\
1. fit to $A_T^{exp}$ under the assumption of $\re(\e_{\ell}) = 0$ and 
$x_{\ell}^{(-)} = 0$ [14], \\
2. fit to $A_{\de}^{exp}$ [15], and \\
3. fit to both $A_T^{exp}$ and $A_{\de}^{exp}$ using as constraints the 
Bell-Steinberger relation and the PDG value of $d_L^{\ell}$ [16], \\
succeeded in determining $\re(\e)$, $\re(\de)$, $\im(\de)$, $\re(\e_{\ell})$, $\im(x_{\ell}^{(+)})$ and/or $\re(x_{\ell}^{(-)})$ simultaneously. 

Among the numerical outputs obtained by CPLEAR Collaboration, there are two 
pieces, $\re(\e) = (1.55 \pm 0.35) \times 10^{-3}$ from [14] and $\re(\de) 
\simeq (0.30 \pm 0.33) \times 10^{-3}$ from [15], are in fact determined 
predominantly by the asymptotic values of $A_T^{exp}(t)$ and $A_{\de}^{exp}(t)$.\footnote{In contrast, the values of $\im(x_{\ell}^{(+)})$, 
$\re(x_{\ell}^{(-)})$ and $\im(\de)$ obtained are sensitive to the behavior of 
$A_T^{exp}(t)$ and of $A_{\de}^{exp}(t)$ at $t$ comparable to $\tau_S$.} One may therefore interpret these outputs as giving the values of $A_T^{exp}(t \gg \tau_S)/4$ and $A_{\de}^{exp}(t \gg \tau_S)/8$ respectively. Replacing Eq.(5.6a) 
with Eq.(7.1a), using Eq.(5.5) instead of Eq.(5.6b), and with the data listed in Table 4 as well as in Table 1 used as inputs, we perform an analysis similar to the previous one [6], and obtain the result shown in Table 5.

A couple of remarks are in order. \\
1. The assumption of $x_{\ell}^{(-)} = 0$ has little influence numerically on 
determination of $\re(\e)$, $\im(\de)$ and $\im(\e+\e_0)$ and the error of these parameters is dominated by that of $\eta_{000}$. \\
2. Our constraint to $\re(\e_{\ell})$ is better to be interpreted as a 
constraint to $\re(\e_{\ell}-x_{\ell}^{(-)})$, the error of which is controlled  dominantly by that of $A_T^{exp}$. \\
3. The error of $\re(\de)$ and $\re(\e_0)$ is also controlled dominantly by that of $A_T^{exp}$. \\
4. The numerical results we have obtained are fairly in agreement with those 
obtained by CPLEAR  Collaboration in [16], except that we have not been able to separate $\re(\e_{\ell})$ from $\re(x_{\ell}^{(-)})$.
%
%%%%%% table 4.

% For tables use
\begin{table}
\caption{Input data (2).}
\label{tab:4}       % Give a unique label
% For LaTeX tables use
\begin{center}
		\begin{tabular}{c|c|c|c|l}
		\hline \hline 
		&Quantity & Value  &Unit & Ref. \\ \hline 

		$3\pi $ &$\ga _L(\pi ^+\pi ^-\pi^0)/\ga _L$ &$12.56\pm 0.20$ &$\% $ & [17] \\ 
		&$\ga _L(\pi ^0\pi ^0\pi^0)/\ga _L$ &$21.12\pm 0.27$ & $\% $ &[17] \\ 
		&$\re (\eta _{+-0})$ &$-0.002\pm 0.008$& & [13,16] \\ 
		&$\im (\eta _{+-0})$ &$-0.002\pm 0.009$ & & [13,16] \\ 
		&$\re (\eta _{000})$ &$0.08\pm 0.11$& & [13,16] \\ 
		&$\im (\eta _{000})$ &$0.07\pm 0.16$ & & [13,16] \\ \hline 

		$\pi ^+\pi ^-\ga $ &$\ga _S(\pi ^+\pi ^-\ga )/\ga _S$ &$0.178\pm 0.005$ &$\% $ & [17] \\ 
		&$|\eta _{+-\ga }|$ &$2.35\pm 0.07$ &$10^{-3}$ & [17] \\ 
		&$\phi _{+-\ga }$ &$44\pm 4$ &$ ^\circ $ & [17] \\ \hline 

		$\pi \ell \nu $ &$\sum_{\ell }\ga _L(\pi \ell \nu )/\ga _L$ &$65.96\pm 0.30$ &$\% $ & [17] \\ 
		&$\im (x_{\ell }^{(+)})$ &$-0.003\pm 0.026$ & & [17] \\ 
		&$A_T^{exp}(t\gg \tau _S)$ &$6.2\pm 1.4$ &$10^{-3}$ &
		 [14] \\ \hline 
		\end{tabular}
\end{center}
% Or use
%\vspace*{5cm}  % with the correct table height
\end{table}

%%%%%% table 5.

% For tables use
\begin{table}
\caption{Constraints (in unit of $10^{-3}$) to $CP$,~$T$ and/or $CPT$-violating prameters (2).}
\label{tab:5}       % Give a unique label
% For LaTeX tables use
\begin{center}
		\begin{tabular}{c|c}
		\hline \hline 
		 Quantity & Result  \\ \hline 

		 $\re(\e) $  & $1.666\pm 0.048$    \\ 
		 $\im (\e + \e _0) $  & $1.590\pm 0.059$  \\ \hline 

		 $\im(\de) $  & $0.020\pm 0.051$    \\ 
		 $\re(\de) $  & $-0.085\pm 0.361$   \\ 
		 $\re(\e_0) $  & $-0.099\pm 0.365$  \\ 
		 $\re(\e_{\ell}) $  & $-0.116\pm 0.353$   \\ \hline 

		\end{tabular}
\end{center}
% Or use
%\vspace*{5cm}  % with the correct table height
\end{table}

%

%%%% Sec.8 %%%%
\section{Case study. -$T$ or $CPT$ violation ?-}

In the analyses given in the previous sections, we have taken account of 
the possibility that any of $CP$,~$T$ and $CPT$ symmetries might violated in 
the $\ko-\kob$ system. Our numerical results shown in Table 3 and Table 5 
indicate that $CPT$ symmetry appears consistent with experiments while $T$ 
symmetry appears not consistent with experiments. To confirm these 
observations, we now go on to perform a case study. \\ \\
{\it Case A. $CPT$ is a good symmetry.}

Putting 
\[
\de = \re(\e_I) = \re(\e_{\ell}) = x_{\ell}^{(-)} = 0~, 
\]%
Eqs.(5.2), (5.5), (5.6a,b), (5.8) and (5.13) reduce respectively to
%
%%% (8.1a,b,c,d,e,f)
%
\begin{subequations}
\begin{equation}
\eta _I \simeq \e + i\im(\e_I)~, 
\end{equation}
\begin{equation}
d_L^{\ell} \simeq 2\re(\e)~, 
\end{equation}
\begin{equation}
d_1^{\ell}(t \gg \tau_S) \simeq 4\re(\e)~, 
\end{equation}
\begin{equation}
d_2^{\ell}(t \gg \tau_S) \simeq 0~, 
\end{equation}
\begin{equation}
d_L^{\ell} \simeq d_1^{\ell}(t \gg \tau_S)/2~, 
\end{equation}
\begin{equation}
\e'/\eta_0 \simeq \re(\omega')\im(\e_2-\e_0)e^{-i\De\phi'}/[|\eta_0|\cos(\de_2-\de_0)]~.
\end{equation}
\end{subequations}
Eq.(8.1f) gives\footnote{It is to be noted that, if and only if $CPT$ symmetry is supplemented with the very accidental empirical fact 
$\phi_{SW} \simeq \de_2-\de_0+\pi/2$, one would have $\im(\e'/\eta_0) \simeq 0$; it is  therefore, as emphasized in [7], not adequate to assume this in a 
phenomenological analysis.}
%
%%% (8.2)
%
\begin{equation}
\im(\e'/\eta_0) = -\re(\e'/\eta_0)\tan \De\phi'~,
\end{equation}
and the simplified version of the Bell-Steinberger relation, 
Eq.(5.17), gives\footnote{Eq.(8.3a) states that deviation of $\phi_0$ from 
$\phi_{SW}$ measures $CPT$ violation. This is equivalent to the more familiar 
statement: deviation of $(2/3)\phi_{+-} + (1/3)\phi_{00}$ from $\phi_{SW}$ 
measures $CPT$ violation, because Eq.(5.9), supplemented with the 
experimental observation $|\eta_{00}/\eta_{+-}| \simeq 1$ and $\De \phi \simeq 0$, gives $\phi_0 \simeq (2/3)\phi_{+-}+(1/3)\phi_{00}$.}
%
%%% (8.3a,b)
%
\begin{subequations}
\begin{eqnarray}
\phi_0 &\simeq& \phi_{SW}~, \\
\re(\e) &\simeq& |\eta_0|\cos\phi_{SW}~.
\end{eqnarray}
\end{subequations}

>From the input data (Table 1 and Table 4) and the intermediate output data 
(Table 2), we observe the following:

(1) The experimental values of $ d_L^{\ell}$, $d_1^{\ell}(t \gg \tau_S)$ and $d_2^{\ell}(t \gg \tau_S)$ are compatible with Eqs.(8.1d,e).

(2) The values of $\re(\e'/\eta_0)$, $\im(\e'/\eta_0)$, $\phi_0$ and $\de_2-\de_0$ are, as illustrated in Fig. 1, compatible with Eq.(8.2).

(3) The values of $\phi_0$ and $\phi_{SW}$ are compatible wiht Eq.(8.3a). 

(4) The values of $\re(\e)$ determined from Eqs.(8.1a) and (8.1b), 
$(1.652 \pm 0.029) \times 10^{-3}$ and $(1.635 \pm 0.060) \times 10^{-3}$, are 
compatible with each other and, as a weighted average, give
%
%%% (8.4)
%
\begin{equation}
\re(\e) \simeq (1.649 \pm 0.026) \times 10^{-3}~,
\end{equation}
which is compatible with $(1.656 \pm 0.014) \times 10^{-3}$ determined with the aid of the Bell-Steinberger relation Eq.(8.3b).\footnote{Eq.(5.15), with $\re(\e_{\ell})=0$, yields $(1.667 \pm 0.048) \times 10^{-3}$.}

(5) Eqs.(8.1a) and (8.1f) give
%
%%% (8.5a,b)
%
\begin{subequations}
\begin{eqnarray}
\im(\e+\e_0) &\simeq& (1.570 \pm 0.030) \times 10^{-3}~, \\
\im(\e_2-\e_0) &\simeq& (3.02~~ \pm 1.09~~) \times 10^{-4}~.
\end{eqnarray}
\end{subequations}
~\\
{\it Case B. $T$ is a good symmetry.}\footnote{The possibility of $CP/CPT$ violation in the framework of $T$ symmetry was examined before by one of the present authors (S.Y.T) [27] when the experimental results which upset $CPT$ symmetry 
(e.g., $|\eta_{00}|$ is nerely twice as large as $|\eta_{+-}|$ !) had been 
reported. The same possibility was recently reconsidered by Bigi and Sanda [24].}

Putting 
\[
\e = \im(\e_I) = \im(x_{\ell}^{(+)}) = \im(x_{\ell}^{(-)}) = 0~, 
\]
Eqs.(5.2), (5.5), (5.6a) and (5.13) reduce respectively to
%
%%% (8.6a,b,c,d)
%
\begin{subequations}
\begin{equation}
\eta _I \simeq -\de + \re(\e_I)~, 
\end{equation}
\begin{equation}
d_L^{\ell} \simeq -2\re(\de)+2\re(\e_{\ell}-x_{\ell}^{(-)})~, 
\end{equation}
\begin{equation}
d_1^{\ell}(t \gg \tau_S) \simeq 2\re(\e_{\ell}-x_{\ell}^{(-)})~, 
\end{equation}
\begin{equation}
\e'/\eta_0 \simeq \re(\omega')\re(\e_2-\e_0)e^{-i(\De\phi'+\pi/2)}/[|\eta_0|\cos(\de_2-\de_0)]~.
\end{equation}
\end{subequations}
Eq.(8.6d) gives 
%
%%% (8.7)
%
\begin{equation}
\im(\e'/\eta_0) = \re(\e'/\eta_0)\cot \De\phi'~,
\end{equation}
and the simplified version of the Bell-Steinberger relation, Eq.(5.17), gives
%
%%% (8.8a,b)
%
\begin{subequations}
\begin{eqnarray}
\phi_0 &\simeq& \phi_{SW} \pm \pi/2, \\
\im(\de) &\simeq& \pm|\eta_0|\cos\phi_{SW}~.
\end{eqnarray}
\end{subequations}

>From the input data (Table 1 and Table 4) and the intermediate output data (Table 2), we observe the following:

(1) As illustrated also in Fig.1, the values of $\re(\e'/\eta_0)$, $\im(\e'/
\eta_0)$, $\phi_0$ and $\de_2-\de_0$ are not compatible with Eq.(8.7). 

(2) The values of $\phi_0$ and $\phi_{SW}$ are not compatible with Eq.(8.8a).

(3) Eq.(8.6a) gives
%
%%% (8.9)
%
\begin{equation}
\im(\de) \simeq (-1.570 \pm 0.030) \times 10^{-3}~,
\end{equation}
to be compared with $\pm(1.656 \pm 0.014) \times 10^{-3}$ determined with the aid of Eq.(8.8b).

(4) Eq.(8.6d) gives
%
%%% (8.10)
%
\begin{equation}
\re(\e_2-\e_0) \simeq (0.61 \pm 3.12) \times 10^{-4}~,
\end{equation}
while Eqs.(8.6a,b,c) give in turn
%
%%% (8.11a,b,c)
%
\begin{subequations}
\begin{eqnarray}
\re(\e_{\ell}-x_{\ell}^{(-)}) &\simeq& (3.14 \pm 1.40) \times 10^{-3}~, \\
\re(\de) &\simeq& (1.51 \pm 1.40) \times 10^{-3}~, \\
\re(\e_0) &\simeq& (3.16 \pm 1.40) \times 10^{-3}~.
\end{eqnarray}
\end{subequations}

The observation (1) establishes the existence of direct $CP/T$ violation in 
the $\ko-\kob$ system [11,12,28].\footnote{We like to mention that Eq.(8.7) 
would  become consistent with experiments if, say, $\phi_{00}$ would prove to 
be away from $\phi_{+-}$ roughly by $\simeq 6^{\circ}$ or more.} The observation (2), though subject to the validity of the Bell-Steinberger relation, also 
implies that $CP/T$ symmetry is violated in the $\ko-\kob$ system.

%%%%%% fig 1.
\vspace*{13pt}
\begin{figure}[htbp] 
\unitlength 0.1in
\begin{picture}(0.06,0.90)(4.00,-52.20)
% \im(\e'/\e)
\put(7.0000,-4.2000){\makebox(0,0)[lb]{$\im(\e'/\eta_0)\times 10^3$}}%
% \re(\e'/\e)
\put(30.0000,-55.2000){\makebox(0,0)[lb]{$\re(\e'/\eta_0)\times 10^3$}}%
\end{picture}
%\vspace*{13pt}
 \centerline{\includegraphics*[width=12cm]{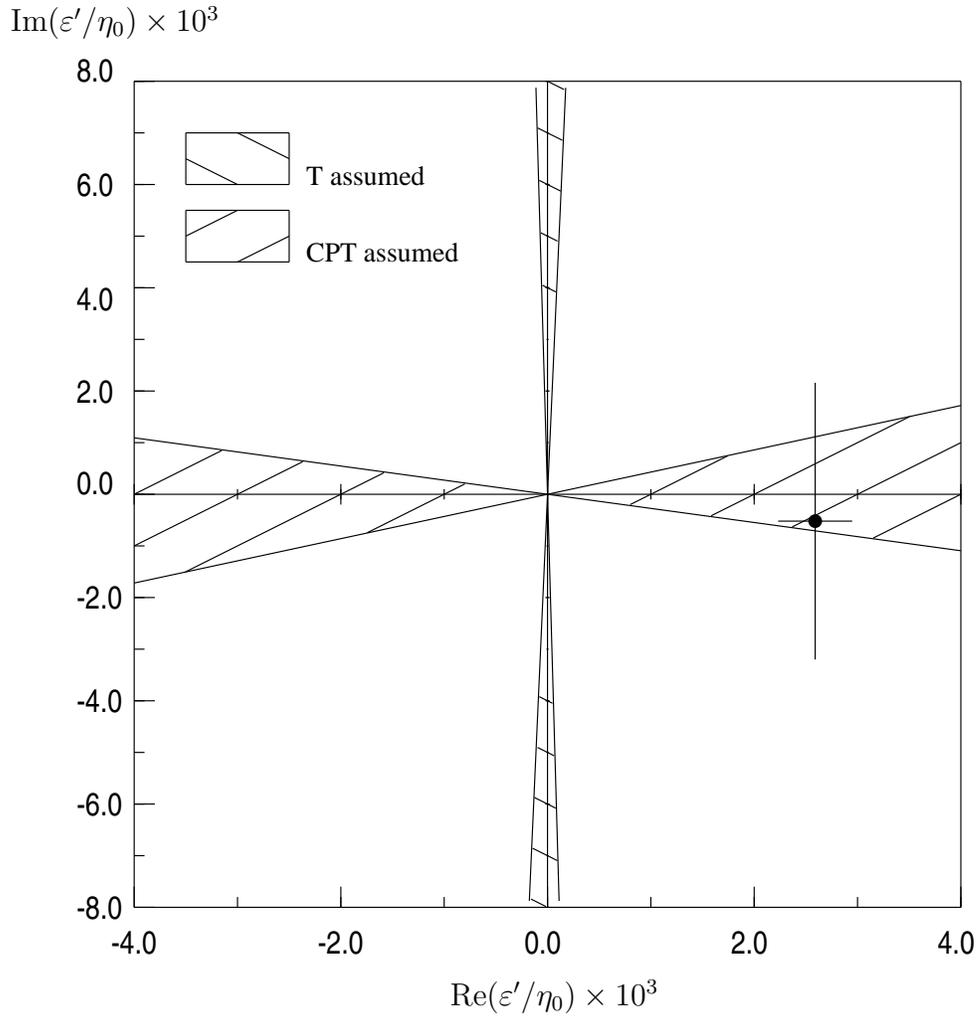}}
\vspace*{10pt}
 \caption{The allowed region on the complex $\e'/\eta_0$ plane when $CPT$ or $T$ symmetry is assumed. (The experimental value of $\e'/\eta_0$ is also shown.)}
\end{figure}
%

%%%% Sec.9 %%%%
\section{Summary and concluding remarks}
In order to identify or search for violation of $CP$,~$T$ and $CPT$ 
symmetries in the $\ko-\kob$ system, parametrizing the mixing parameters and the relevant decay amplitudes in a convenient and well-defined way, we have, with 
the aid of the Bell-Steinberger relation and with all the relevant experimental data used  as inputs, performed numerical analyses to derive constraints to 
the symmetry-violating parameters in several ways. The analysis given in Sec.6 
is based on the data on $2\pi$ decays as well as the well measured leptonic 
asymmetry $d_L^{\ell}$, while, in the analysis given in Sec.7, the data on 
$3\pi$ and $\pi^+\pi^-\ga$ decays and on the newly measured leptonic asymmetries are also taken into account.

The numerical outputs of our analyses are shown in Table 3 and Table 5, and the main results may be summarized as follows:

(1) The $2\pi$ data directly give $\im(\e_2-\e_0) = (2.95 \pm 1.13) \times 
10^{-4}$ in general, or $(3.02 \pm 1.09) \times 10^{-4}$ if $CPT$ symmetry is 
assumed, where possible large uncertainty associated with $\de_2-\de_0$ has been fully taken into account. This result indicates that $CP$ and $T$ symmetries 
are definitively violated in decays of $\ko$ and $\kob$ into $2\pi$ states.

(2) The well-measured leptonic asymmetry $d_L^{\ell}$ directly gives $\re(\e-\de+\e_{\ell}-x_{\ell}^{(-)})=(1.635 \pm 0.060) \times 10^{-3}$, which implies presumably 
that $CP$ and $T$ violations are present also in the the $\ko-\kob$ 
mixing (i.e., $\re(\e) \neq 0$).\footnote{Of course, $d_L^{\ell} \neq 0$ does 
not exclude $CPT$ violation (i.e., $\re(\de-\e_{\ell}+x_{\ell}^{(-)}) \neq 0$.)}

(3) The Bell-Steinberger relation, with the $2\pi$ intermediate states alone 
taken into account, gives $\re(\e) = (1.656 \pm 0.014) \times 10^{-3}$ and 
$\im(\e+\e_0)=(1.566 \pm 0.014) \times 10^{-3}$. If $CPT$ symmetry is assumed, 
$\re(\e)$ is determined without recourse to the Bell-Steinberger relation to be $(1.649 \pm 0.026) \times 10^{-3}$. All these indicate that $CP$ and $T$ 
violations are present in the mixing parameters.

(4) The parameters, nonvanishing of which signals $CP$ and $CPT$ violations, 
have also been constrained. The $2\pi$ data alone directly give $\re(\e_2-\e_0) = (0.084 \pm 0.328) \times 10^{-3}$ and, with the aid of the Bell-Steinberger 
relation, give $\im(\de) = (-0.004 \pm 0.027) \times 10^{-3}$, 
$\re(\de - \e_0) = (0.004 \pm 0.026) \times 10^{-3}$ and $\re(\de - \e_{\ell} +
 x_{\ell}^{(-)}) = (0.021 \pm 0.062) \times 10^{-3}$. These results imply that 
 there is no evidence for $CPT$ violation on the one hand and that $CPT$ 
symmetry is tested at best to the level of a few $\times 10^{-5}$ on the other 
hand.
 
(5) The Bell-Steinberger relation, even with the intermediate states other than the $2\pi$ states taken into account, still allows one to determine $\re(\e)$ 
and $\im(\e+\e_0)$ and to constrain $\im(\de)$ to a level better than $10^{-4}$. On the other hand, the constraint to $\re(\de)$, $\re(\e_0)$ and $\re(\e_{\ell}-x_{\ell}^{(-)})$ is a little loose and is at the level of a few $\times 10^{-4}$.

The recent data reported by KTeV Collabotration [11] and NA48 Collaboration [12] are extremely remarkable in that they play a vital role in establishing $\im(\e_2-\e_0) \neq 0$, and that this is at present the only piece which indicates 
"direct violation" (in the sense defined in Sec.4) of $CP$ and $T$ symmetries 
and thereby unambiguously rules out superweak (or superweak-like) models of $CP$ violation.

The analyses done by CPLEAR Collaboration [14-16] are also very remarkable in 
particular in that they have succeeded in deriving constraint to $\re(x_{\ell}^{(-)})$, and in that they have determined $\re(\e+\e_{\ell}-x_{\ell}^{(-)})$ and $\re(\de)$ directly (i.e., without invoking the Bell-Steinberger relation) with accuracy down to the level of a few $\times 10^{-4}$.\footnote{Alvarez-Gaume, Kounnas and Lola [29] further claim that the CPLEAR results allow one to conclude, without invoking Bell-Steinberger relation, that $T$ symmetry is violated 
independent of whether $CP$ and/or $CPT$ symmetries are violated or not.}

It is expected that the new experiments at the facilities such as DA$\Phi$NE, 
Frascati, will be providing data with such precision and quality that a more precise and thorough test of $CP$, $T$ and $CPT$ symmetries, and a test of the 
Bell-Steinberger relation as well, become possible [23,30,31].
%
%%%%%% References %%%%%%
%

\end{document}